\documentclass[leter,twocolumn,11pt]{article} 
\usepackage{flushend} 
\usepackage{amssymb}
\usepackage{xcolor}
\usepackage{ae}
\usepackage[all]{xy}
\usepackage{tikz}
\usepackage{graphicx}
\usetikzlibrary{backgrounds,calc}
\usepackage{amsmath}
\usepackage[top=0.8in, bottom=0.8in, left=0.6in, right=0.6in]{geometry} 
\usepackage{url}
\usepackage{hyperref}
\usepackage{comment} 
\usepackage{cite}    

\begin{document}
\sloppy 
\renewcommand{\tablename}{Tabla}
\renewcommand{\figurename}{Fig.}

\title{
        {\LARGE Galileo's Quantization}
}

\author
{ 
Paco H. Talero L.$^{1,2\, *}$ $\:$ and $\:$ Inés Delgado R.$^{2,3\, \dagger}$ \\  
\small $^{1}$ Facultad de Ingeniería y Ciencias Básicas, Universidad Central, Bogotá D.C. Colombia \\
\small $^{2}$ Facultad de Ciencias y Educación, Universidad Distrital Francisco José de Caldas-PCLF, Bogotá D.C. Colombia \\
\small $^{3}$ Secretaría de Educación del Distrito, Bogotá D.C. Colombia \\
\small \textcolor{blue}{\textit{ptalerol@ucentral.edu.co}}$^{*}$, \textcolor{blue}{\textit{idelgador@udistrital.edu.co}}$^{\dagger}$
}
\date{\today}
\twocolumn
[
\begin{@twocolumnfalse}
\maketitle 
\begin{abstract}
In the context of physics didactics, alternative instructional approaches have often been employed to facilitate conceptual understanding of various topics. In this article, an alternative formulation for analyzing the motion of bodies on inclined planes is presented, based on Galileo's experimental rule involving multiples of odd numbers. According to this, bodies traverse distances that increase in odd multiples of an initial amount during successive equal time intervals. In the proposed formulation, a function is established between distance, whether ascending or descending, and time, omitting the use of Newtonian concepts such as velocity, acceleration, or force. The theoretical exploration of this formulation may provide significant didactical benefits, offering scenarios for discussion and reflection in physics didactics.\\

\textbf{Keywords:} physics didactics, Galileo's odd-number rule, inclined plane, alternative formulation. \\

%

\end{abstract}
\end{@twocolumnfalse}
]

\section{Introduction}

The importance of alternative formulations to well-established theories in various fields of physics has proven fundamental to the advancement of science. Notable examples include Lagrangian, Hamiltonian, and Newtonian mechanics (NM), as well as the path-integral formulation of quantum mechanics \cite{landau1976mechanics,goldstein2002,feynman_path}. In the educational field, the geometric formulation described in \cite{feynman_lost} provides a geometric approach to planetary motion, offering a distinct and enriching perspective. Another example is the algebraic proof presented in \cite{geoFermat}, which connects Fermat's principle with Snell's law using only algebraic reasoning. This alternative deduction avoids calculus and serves as a useful tool for introductory physics courses. \\

In this context, within the framework of physics teaching and as an example showcasing the importance of history in physics education \cite{decamp_implementing}, this article aims to present an alternative formulation of the motion of bodies on inclined planes, based on Galileo’s work. This approach emphasizes that, for a given time, the distances traveled in successive time intervals are quantized, this means that the body moves in distances that are $3$, $5$, $7$, and so on, times the initial distance traveled \cite{galilei1638}. This formulation is used to explain the distance-time relationship in body motion without relying on kinematic or dynamic concepts such as velocity, acceleration, gravity, force, energy, and so on.\\

In this way, using Galileo's quantization (GQ) postulate, the postulate of reversibility, and only elementary algebra, a quadratic distance-time relationship was derived for a body moving up or down an inclined plane, which is consistent with those obtained from NM.\\

In Sec.$2$, the methodology used to formulate GQ theory is explained. Subsequently, in Sec.$3$ the theory is formulated. In Sec.$4$, the theoretical and experimental context is discussed in relation to didactic perspectives. Finally, in Sec.$5$ the conclusions are presented.
 
\section{Methodology}

To understand how GQ theory will be formulated in this article, a short definition of physical theory will need to be established, which is done in this section.\\

A physical theory is a systematic representation of a set of natural phenomena, formulated in precise language and, when necessary, with the use of logical-mathematical formalism. To fulfill its function, it exhibits internal coherence, meaning a logical structure free of contradictions within its concepts and postulates. Furthermore, it establishes an explanatory and predictive correspondence with observable facts, enabling it not only to describe phenomena but also to anticipate results under specific conditions \cite{bunge_philosophy}. The theory must have a well-defined range of reference that clearly delimits the set of phenomena or situations it is intended to represent, and its formalism must be closed under the operation of deduction \cite{Romero2021scientific}.\\

In general terms, it should align with the results accepted in other scientific fields, considered true up to the present. This link to reality is subject to constant revision, allowing the theory to be refined and adjusted in light of new data and scientific advancements \cite{Romero2021scientific,feynman_character}.\\

In line with the above, a theory does not need to have explanatory or predictive scope over a wide range of facts; it is sufficient for it to cover only a few. Although less comprehensive, such a theory may still prove useful for didactic purposes.\\

\section{Results}

In this section, GQ theory is presented, including its postulates, fundamental theorems, and overarching considerations.

\subsection{Galileo's theory of quantization}

Galileo discovered a natural law describing the motion of bodies descending on inclined planes; specifically, it states that these bodies cover distances in a sequence of odd numbers from the initial distance during each equal time interval. The theory of GQ proposed here takes this law as one of its postulates, relating the distance traveled by a body on an inclined plane to the time taken for that movement. This theory is presented using postulates and is accompanied by theorems that define its scope and limitations, without the need to resort to concepts from NM.

\subsubsection{Ontological reference of the theory}

Here, the specific objects, processes, or entities in reality that the Galileo quantization theory seeks to represent are specified.  In other words, it refers to the set of entities in the world that act as reference points for the propositions of this theory. This concept pertains to the elements in the world that provide meaning to the conceptual structure of this theory \cite{bunge_philosophy,Romero2021scientific}.\\

The objective of this theory is to establish a relationship between the distance traveled in a straight line by a body moving on an inclined plane, whether descending or ascending, and the time taken for that movement.\\

The inclined plane along which the body moves mustn't deform; therefore, it must be constructed from firm wood or metal. Additionally, to ensure the body follows a straight-line trajectory without deviations or jumps, the surface must be flat and include guide channels to further guarantee rectilinear movement.\\

The bodies that move on the inclined plane must satisfy the following characteristics: they must be made of metal or glass to reduce the influence of air or the atmospheric medium; must have a defined shape, such as spheres, cylinders, rings, or parallelepipeds, among others, to prevent jumps and deviations from a straight path; and must have dimensions much smaller than the length of the plane.\\

Finally, the entire system must be located on the surface of the Earth.

\subsubsection{Postulate 1: Quantization} \label{pos1}

If a body is released at the top of an inclined plane and descends along it, it does so in such a way that it travels a distance \(d\) in the first time interval \(\tau\), in a second equal interval, it descends \(3d\), in the third \(5d\), in the fourth \(7d\), and so on. Here $d$ and $\tau$ are derived from experience; if \(d\) is established, \(\tau\) is measured, and vice versa. The above depends fundamentally on the angle of inclination of the plane and the location where it is situated, such as Earth, the Moon, or other places.\\

The concept of quantification expressed in this postulate arises from the observation that, regardless of the values of \(\tau\) and the distance \(d\) it represents, for times that are multiples of \(\tau\), the distance traveled by the body on the plane is not arbitrary but instead an odd multiple of the distance traveled during the first interval. However, it is important to note that this concept is in no way connected to quantum mechanics.

\subsubsection{Postulate 2: Reversibility} \label{pos2}

Some natural laws are independent of the direction of time \cite{feynman1963feynman}. If a body descends an inclined plane, the reverse movement (ascent) may exist. This conjectural idea forms the basis of the second postulate.\\

Second postulate: the descending motion of a body on an inclined plane has an inverse motion that starts at the bottom of the plane and ends at the top, reproducing the descent in reverse temporal sequence. For example, if in the first time interval \(\tau\) the body ascends a distance of \(7d\), then in the next interval \(\tau\) it covers \(5d\), later \(3d\), and so on until it comes to a stop.

\subsubsection{Theorem 1: The quadratic relationship between distance and time of descent}

According to postulate \ref{pos1}, the distance \( D_l \) traveled by the body over \( n \) time intervals as it descends the inclined plane is given by the sum of the first \( n \) odd numbers, multiplied by the distance covered during the first time interval. Specifically

\[
D_l = \left[ 1 + 3 + 5 + 7 + 9 + 11 + \cdots + (2n-1) \right] \: d.
\]

It is important to note that this sum is equal to \( n^2 \) \cite{swokowski2002algebra}. Therefore, the total distance covered up to the \(n\)-th time interval can be expressed as

\begin{equation} \label{eq:distancia}
D_l = n^2 d.
\end{equation}

On the other hand, the elapsed time \( t \) is given by \( t = n \tau \), which implies that
\begin{equation} \label{eq:time}
n = \frac{t}{\tau}.
\end{equation}

Substituting \( n \) from Eq.(\ref{eq:time}) into Eq.(\ref{eq:distancia}), the result is 

\begin{equation} \label{eq:distancia_t}
D_l = \eta t^2,
\end{equation}

with \( \eta = \frac{d}{\tau^2} \), which is derived from experience and is also constant.\\

Furthermore, if \( \eta \) is small, then \( D_l \) can cover distances small enough to allow any experimental comparison. Thus, the result in (\ref{eq:distancia_t}) is comparable to that obtained in textbooks through NM; see, e.g., \cite{Arons}.

\subsubsection{Theorem 2: Quadratic relationship between distance and time in ascent}

Let \( L \) be the length of the plane on which the body moves, \( N \) the maximum number of length intervals, and \( q \) the maximum odd number, where \( q = 2N - 1 \). Thus, from Eq.(\ref{eq:distancia}), it is obtained that

\[
L = N^2 d   \quad \text{and therefore} \quad      d = \frac{L}{N^2}.
\]

From postulate \ref{pos1}, it follows that the first distance is \( qd \). Let \( z = qd \); and combining these equations, it is obtained that:
\begin{equation} \label{ult}
z = \frac{q}{N^2} L
\end{equation}

Therefore, when $\tau$ is defined, the body traverses a distance $d$ during descent in the first time interval and a distance $z$ during ascent, also within the first time interval.\\

Let $D_{u}$ denote the total distance traveled upwards by the body upon completing the $k$-th time interval, where $k \leq N$. Additionally, let $S_r$ represent the distance traveled during each interval $r = 1, 2, 3, \dots, k$, with $S_1 = z$. Thus,
\[
S_r=\left [q-2\left ( r-1 \right ) \right ]\,z
\]
when added together, it is found
\[
D_u=\sum_{r=1}^{k}\left [q-2\left ( r-1 \right ) \right ]\,z
\]

and expanding the terms, it is obtained
\[
D_u=\left [ qk-k\left ( k+1 \right )+2k \right ]\,z
\]

When the terms are expanded and $k= \frac{t}{\tau}$ is used, the result obtained is

\[
D_u=\left ( 1+q \right )\left ( \frac{z}{\tau } \right )t+   \left( \frac{z}{\tau^2} \right)\, t^2
\]

making the following substitutions

\[
\alpha = \frac{(1+q)z}{\tau} \quad \text{and} \quad \beta = \frac{z}{\tau^2}, 
\]
finally, it is found

\begin{equation} \label{qua_sub}
D_{u} = \alpha \, t - \beta \, t^2. 
\end{equation}

The Eq.\eqref{qua_sub} represents the function describing the distance climbed by the body on the inclined plane as a function of the elapsed time. This result aligns with those presented in texts that derive the motion from NM; see, e.g., \cite{Arons}.

\section{Discussion}

The GQ theory, defined in this work, is regarded as a valuable theoretical tool for teaching introductory physics. This approach offers an innovative analysis of the motion of bodies on inclined planes, concentrating solely on the relationship between the distance traveled and the time elapsed, without relying on NM or the traditional geometric perspective historically used.\\

This theory is expected to provide an alternative theoretical perspective for teaching uniformly varied motion (UVM), targeting high school students and first-semester students in sciences and engineering. Furthermore, it is expected to serve as a valuable tool for exploring key principles of the epistemology of science, as GQ allows the illustration of concepts such as the postulates and limitations of a theory, regularities in nature, the relationship between theory and experiment, the importance of rigorous quantification, and a theory's internal coherence, among other foundational aspects.\\

The use of algebra, instead of Galileo's original geometric approach, is expected to enable students to analyze motion on inclined planes and internalize the concept of UVM without employing NM principles. Additionally, by comparing the perspectives of GQ with those of NM, students are expected to strengthen their critical thinking, broaden their theoretical understanding, stimulate creativity, and consolidate their mastery of physical concepts.\\

On the other hand, in laboratory practices where the theoretical perspective of GQ is applied, it has been observed that students assume distances within time intervals \(\tau\) are divided into equal segments, contradicting the first postulate of the GQ theory. Furthermore, they often apply the rule of three automatically, highlighting a cognitive difficulty in disengaging from the frameworks of NM and classical kinematics when explaining the motion of bodies on inclined planes, whether moving up or down. Physics education researchers are encouraged to explore the cognitive and didactic aspects of these challenges more deeply, aiming to strengthen scientific and physical reasoning while designing more effective strategies to address them.

\section{Conclusions}
In this work, a simple theory has been formulated to establish a relationship between the distance traveled and the elapsed time for bodies moving up or down inclined planes, without relying on the concepts of NM.  This theory has practical applications in teaching UVM, offering an innovative approach. Although it lacks the predictive and explanatory power of NM, its significance lies in its effectiveness as a teaching tool. Future research is encouraged to explore the didactic and cognitive contributions that emerge when students attempt to understand how it is possible to establish a distance-time relationship in the motion of bodies moving up or down inclined planes without relying on the concepts of acceleration and velocity


\end{document}